# String Matching with Multicore CPUs: Performing Better with the Aho-Corasick Algorithm

S. Arudchutha, T. Nishanthy and R.G. Ragel
Department of Computer Engineering
University of Peradeniya, Sri Lanka

*Abstract -* **Multiple string matching is known as locating all the occurrences of a given number of patterns in an arbitrary string. It is used in bio-computing applications where the algorithms are commonly used for retrieval of information such as sequence analysis and gene/protein identification. Extremely large amount of data in the form of strings has to be processed in such bio-computing applications. Therefore, improving the performance of multiple string matching algorithms is always desirable. Multicore architectures are capable of providing better performance by parallelizing the multiple string matching algorithms. The Aho-Corasick algorithm is the one that is commonly used in exact multiple string matching algorithms. The focus of this paper is the acceleration of Aho-Corasick algorithm through a multicore CPU based software implementation. Through our implementation and evaluation of results, we prove that our method performs better compared to the state of the art.**

## I. INTRODUCTION

Multiple string matching algorithms are used for finding all occurrences of a set of patterns in an arbitrary string. It is commonly used in information retrieval, text editing applications, spam filters and virus scanners. There are several applications for multiple strings matching process in computational biology like protein sequence and DNA synthesis [15]. The techniques used for identifying the newly sequenced DNA or protein generally involve comparing the sequence of known organisms to the newly sequenced organism. Since the data in the form of strings in the biological database is growing at an exponential rate, there is always a need for improving the performance of multiple strings matching algorithms.

With the development of technology, desktop and laptop computer architectures have evolved into multicore processors. Such processors generally provide vast computing capabilities, but they require parallelizing the existing applications in order to get advantage of the existing hardware. Consequently, applications and algorithms must be modified and adapted to get advantage of all available resources of a modern computer.

Many researches have been done utilizing both hardware and software to accelerate string matching in several areas: hardware supported approaches use FPGA [7], [12], GPU [3], [8], [9] and Cell/B.E. processor [5] and software based approaches use multiple processors [2], [4]. Among them, the software based acceleration techniques need only some modification in the software code or the architecture. Since the most commonly used multiple string matching algorithm is Aho-Corasick, many attempts were made to accelerate the Aho-Corasick algorithm.

As the Aho-Corasick algorithm is already implemented on multicore processor by another research group [2], our focus is to improve the software implementation further to accelerate Aho-Corasick and to outperform the work presented in [2]. We achieve our objective by taking a pattern focused approach as opposed to an input focused approach, details of which are discussed later in this paper.

The remaining part of the paper is organized as follows: Section II contains the background of the Aho-Corasick algorithm and in Section III, we present related work. Our implementation details and the results of our work are given in Sections IV and V respectively. Finally we conclude the paper in Section VI.

## II. BACKGROUND

We have explored that there are two popular and different implementations of the Aho-Corasick algorithm: (1) Parallel Failure-less Aho-Corasick Algorithm [3] and (2) Aho-Corasick Algorithm with Failure Links [1]. In both implementations, the algorithm consists of the following two steps:
1) Constructing a string-matching machine (a finite state machine) for a given set of patterns.
2) Processing the input string using the string matching machine in a single pass.

However, in both of the implementations, the methods used for constructing the string-matching machine and the method used for processing the input string are different. Let us look at each implementation in detail.

*A. The Aho-Corasick Algorithm with Failure Links*

Fig. 1 is the Aho-Corasick (AC) state machine of the given four patterns, "AB", "ABG", "BEDE", and "EF". Final states of each pattern are represented by double circle. In Fig. 1, the valid transitions are represented by solid lines whereas the failure transitions are represented by dotted lines.

The purpose of the failure transitions is back-tracking the state machine to identify patterns in different locations. Given an input character and a current state, first the AC state machine checks for a valid transition for that input character; if there is no valid transition in that state, the machine jumps to the next state where the failure transition points to. Then, the string-matching machine considers the same character input until the character finds a valid transition. During the transition, it is in the final state of any given pattern then it produces the corresponding pattern as the output.

For example, take an arbitrary input stream with a substring "ABEDE" [3]. First the AC state machine starts from state 0 and traverses to state 1 and then to state 2. Since that is the final state for pattern "AB", the machine produces pattern "AB". Because state 2 does not have a valid transition for the character "E", the state machine then takes a failure transition to state 4 and then considers the same character of input stream (that is "E") and goes to state 5. The AC state machine finally arrives at state 7, which is the final state of pattern "BEDE" and produces the pattern "BEDE".

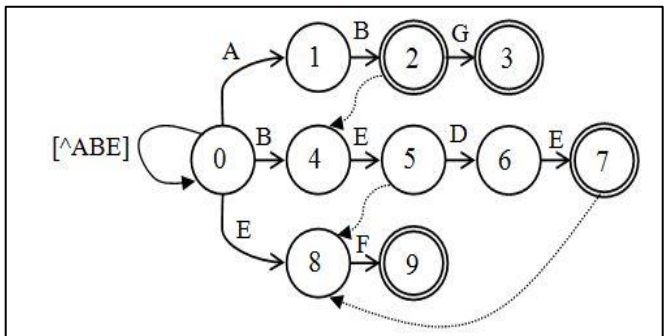

Fig.1. Aho-Corasick state machine for patterns "AB","BEDE","ABG" and "EF"

### B. Parallel Failure-less Aho-Corasick Algorithm

In Parallel Failure less Aho-Corasick (PFAC) implementation, all failure transitions are removed from the state machine. Fig. 2 is the state machine for the PFAC implementation for the same four patterns, "AB", "ABG", "BEDE" and "EF" that we used in the last example.

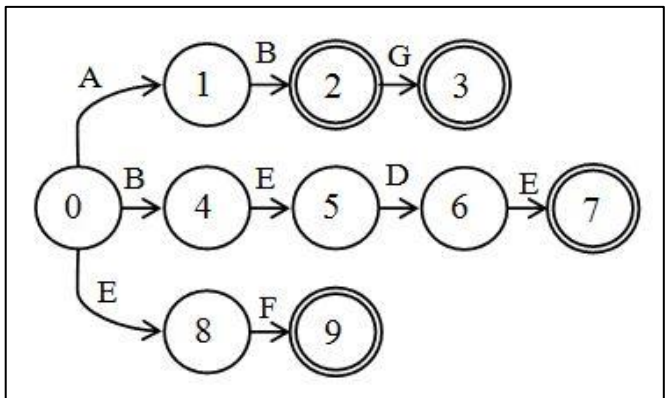

Fig.2. Failure less state machine for patterns "AB", "ABG", "BEDE", and "EF".

PFAC implementation is originally proposed as the GPU version of the Aho-Corasick algorithm. In PFAC, each thread is allocated to a character of the input stream to recognize any pattern starting from the thread starting character. Fig. 3 shows how the PFAC implementation will allocate each character of an input stream a thread to traverse the same state machine in parallel.

The Aho-Corasick state machine with failure links uses the failure transitions for back-tracking the state machine and therefore to identify the patterns that are starting at any character of an input string. However, in the PFAC algorithm, a thread considers only the patterns starting at a particular character and therefore the threads of PFAC do not need to back-track the state machine. For that reason, all the failure transitions of the state machine are removed in the PFAC implementation.

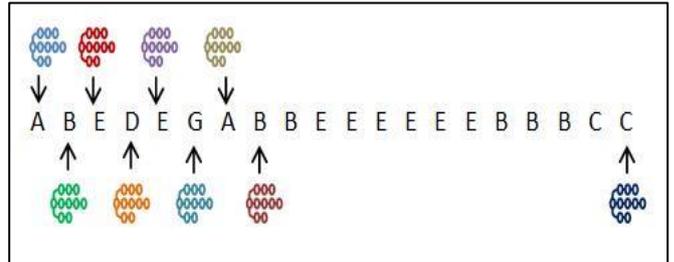

Fig.3. PFAC algorithm each character of an input stream is allocated a thread to traverse the state machine.

As before, consider any input stream with a substring "ABEDE". Let us assume that thread $t_n$ is assigned to character "A" as shown in Fig. 4 to traverse the failure less state machine. After traversing the substring "AB", thread $t_n$ arrives state 2, where matched pattern "AB" is produced. Because state 2 does not have a valid transition for character "E", thread $t_n$ terminates at state 2. Similarly, thread $t_{n+1}$ is assigned to input character "B". After traversing substring "BEDE", thread $t_{n+1}$ arrives state 7 that indicates the matched pattern "BEDE".

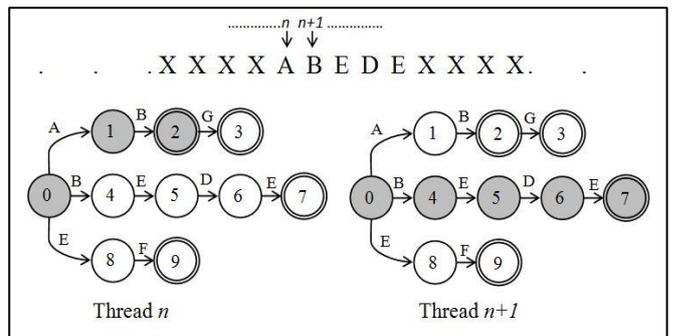

Fig.4. Example of PFAC

### III. RELATED WORK

As it is necessary to find more effective string matching algorithms and their implementations, an appreciable number of researches have been being carried out in this area. Here we present some research work performed on string matching algorithms.

One of the researches [2] done was to examine the hypothesis that in this kind of computation multicore architectures should provide scalable and better performance, but still it would depend on the programming model being utilized as well as on the algorithm selected. The researchers in [2] have explored the two possible different architectures for parallel execution:
1. Splitting the input strings into smaller chunks and processing them in separate threads for pattern matching.
2. Splitting the pattern-set and in separate threads creating different pattern machines and passing the same input string to each different patter machine.

Because of the fact that in classic bio-computing applications, the pattern text would be quite large compared to the input string, the time consumed for creating the state machine would reasonably effect the total time taken for pattern matching. Therefore, the authors in [2] took the second approach where the large pattern-set is divided into smaller chunks. Then, each pattern matching machine with failure link was developed by a different thread and at runtime each machine processed the same input string individually. The throughput was calculated as the number of patterns found per second and this experiment was repeated for different the number of threads being used. Their implementation resulted in 9x throughput improvement on an 8-core processor (that supports 16-threads) compared to a single thread implementation.

In one of the researches done on GPU [3], the authors have increased the throughput by increasing parallelism with GPU. In their method, because they are removing the failure transitions, they do not need to back-track the state machine; hence their implementation is reducing the complexity of the algorithm and the memory usage. They allocated each character of an input string to a GPU thread. Therefore, all the threads of PFAC traverse the same failure less Aho-Corasick state machine and terminates if there is no valid transition.

The classic hardware-based acceleration techniques either need special hardware like general purpose graphics processing units (GPGPUs) [9] or need to build a new hardware like an FPGA based design [7].

One other paper presents a software based implementation of the Aho-Corasick (AC) algorithm and was tested using dictionary on the Cray XMT multithreaded shared memory architecture [4]. They used an improved version of AC algorithm called AC-opt where its state machine is obtained from original by replacing all the failure transitions with regular ones. Because, one research group [10] implemented (Knuth-Morris-Pratt, Boyer-Moore, Aho-Corasick, and Commentz-Walter algorithms directly from the abstract algorithms derived and presented in [11]) and analysed the performance of algorithms without considering the pre-computation time. From the result they have proved that the AC-opt algorithm performs better compared to the original AC.

Focusing on Network Intrusion Detection Systems (NIDS), on the Cell/B.E processor, a research was done on high-speed string searching against large dictionaries [5]. Their aim is performing the exact string matching against large dictionaries by parallelizing AC on the IBM Cell/B.E processor using an improved implementation. Their algorithm has been implemented in C language using the CBE language extension and intrinsic.

In [6], the authors compare the performance of several software based implementations of the AC algorithm on shared-memory architectures and distributed memory architectures using dictionaries by considering the input focussed approach.

As noted in [2], the pattern texts are usually quite large compared to the input text in bio-computing applications and therefore the time consumed for creating the state machine would considerably effect the total time consumed. Furthermore, in PFAC implementation [3] the group got better performance on GPUs. Therefore, to optimize the total time by considering the time taken to build the pattern matching machine and to prove our hypothesis that in PFAC implementation time taken to build the machine is much less than the failure link implementation due to the removal of failure transitions, we are proposing a different method on CPU using parallel failure less AC algorithm techniques. In short, our technique adapts the PFAC algorithm into a multicore CPU implementation.

IV. DESIGN AND IMPLEMENTATION

Here, we give our design and implementation details.

A. Thread Assignment Methodology

Because of the limited number of threads available on CPUs, we cannot allocate each character of an input stream a thread as done in PFAC implementation. Therefore, as shown in Fig. 5, firstly the pattern-set was divided into a given number of threads and then the same number of threads will be created. Then the different pattern chunks are passed onto each thread. It will create different failure less pattern machine.

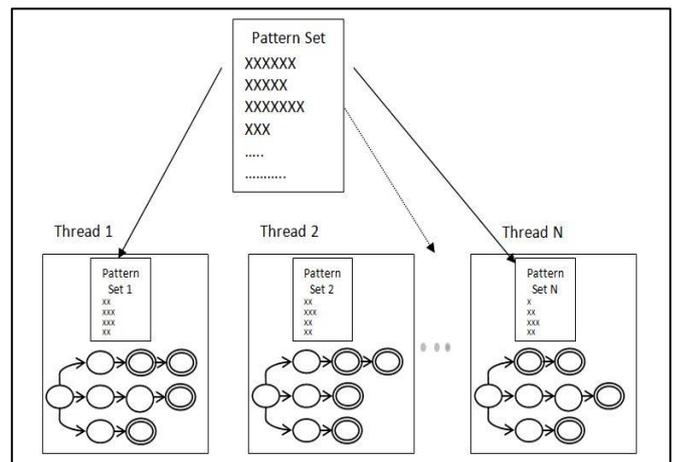

Fig.5. Thread based implementation

We are going to allocate the first character of an input string for each thread. If we consider one thread, when it terminates, it will be allocated to the second character of the input string. Likewise each thread identifies any pattern starting from the starting locations of the input string. For example, consider the input stream "USHERS" and consider one thread that has the pattern-set "HE", "HIS", "SHE" and "HERS". Fig. 6 shows the state machine of the thread mentioned.

Now considering the state machine we mentioned in the last paragraph, first it takes the input from the first character of the input stream. Because state 0 do not have the valid transition for the character "U", it will take the input from the second character of the input stream. Therefore, for character "S" it traverses to state 5 from state 0. After two more traversals, finally the state machine reaches state 7. There it produces the pattern "SHE". For character "R", in state 7 there is no valid transition and therefore it will take the input from the third character of the input string. It leads to state 2 for pattern "HE" and finally reaches state 9 for pattern "HERS". Next, the machine will take the input from the fourth byte of the input stream. Likewise it will consider the inputs until the last character of the input stream.

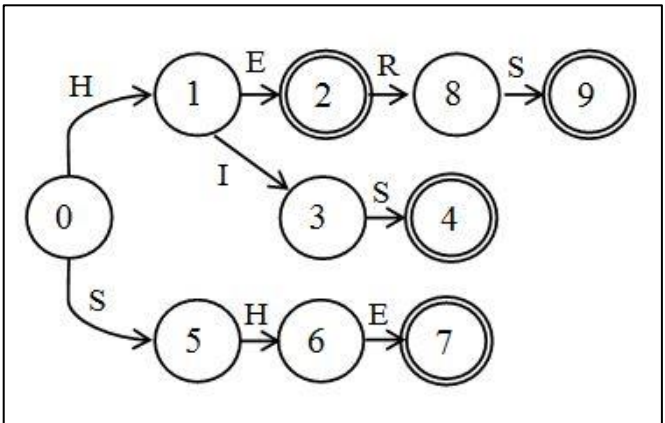

Fig.6. Failure less state machine of patterns "HE", "HIS", "SHE" and "HERS"

*B. Implementation Details*

Initially, we found an acceptable implementation for the Aho-Corasick algorithm. We used "Multifast-v0.6.2" [13] module as the starting point for our implementation. It provides an understandable codebase and a clear implementation. Basically, multifast takes a set of finite pattern strings as an array and an input string and outputs the details on the patterns matched such as their positions in the input string. Furthermore, multifast takes both the input string and the set of patterns from reading the files hence both set of strings can be given as files. We took the following steps to achieve our design: Multifast-v0.6.2 contains failure link transitions in state machines for back-tracking patterns starting at different locations. In our design, we rebuilt the state machines without the failure link transitions and as shown in Algorithms 1 and 2 we changed the searching method with respect to the failure less transitions.

**Algorithm 1.** Searching method in Aho-Corasick.
**Input**. Input file
**Output**. Matched patterns
**Method**. The procedure fread() reads the input and inserts into the buffer. And the buffer is passed into the search method. This will continue until end of the file.

    **begin**
      **do**
        num_read ← fread(*buffer)
        ac_automata_search(&buffer)
    **while** num_read >= read_element_num
    **end**

**Algorithm 2.** Searching method in our implementation
**Input**. Input file
**Output**. Matched patterns
**Method**. After first read we read the input character by character.

    **begin**
      num_read ← fread(*buffer);
      **do**
        ac_automata_search(&buffer);
        **for** i ← 0 **until** buffer.length-1
          **do** buffer[i]=buffer[i+1];
        num_read_tmp ← fread(&c);
        **if** num_read_tmp==1
          **do** buffer[i]=c;
        **if** num_read_tmp==0
          **do** buffer.length—;
      **while** buffer.length>1
    **end**

We implemented parallelism by using the Pthreads [14] library. Pthread is a C language threads programming interface for UNIX that was standardized by IEEE POSIX 1003.1003.1c standard. Pthread allows creating a new thread within the caller process. We used C programming language and for thread handling, POSIX Pthread library was used. All our implementation was carried out on GNU/Linux platform.

## V. EXPERIMENTAL EVALUATION

We analysed the result mainly under the following aspects:

1. Throughput: the number of bytes being processed per unit time.

$$\text{Throughput} = \frac{\text{No. of bytes in the input}}{\text{Total time for processing}} \quad (1)$$

2. Scalability: the effect of changing the input string size and the pattern-set size used in our implementation.

When we used two same size input strings one with several patterns embedded and the other with no patterns embedded,

the time taken for searching did not change significantly. It assured us that the searching time largely depends on the input string size and not on how many patterns found in the input string.

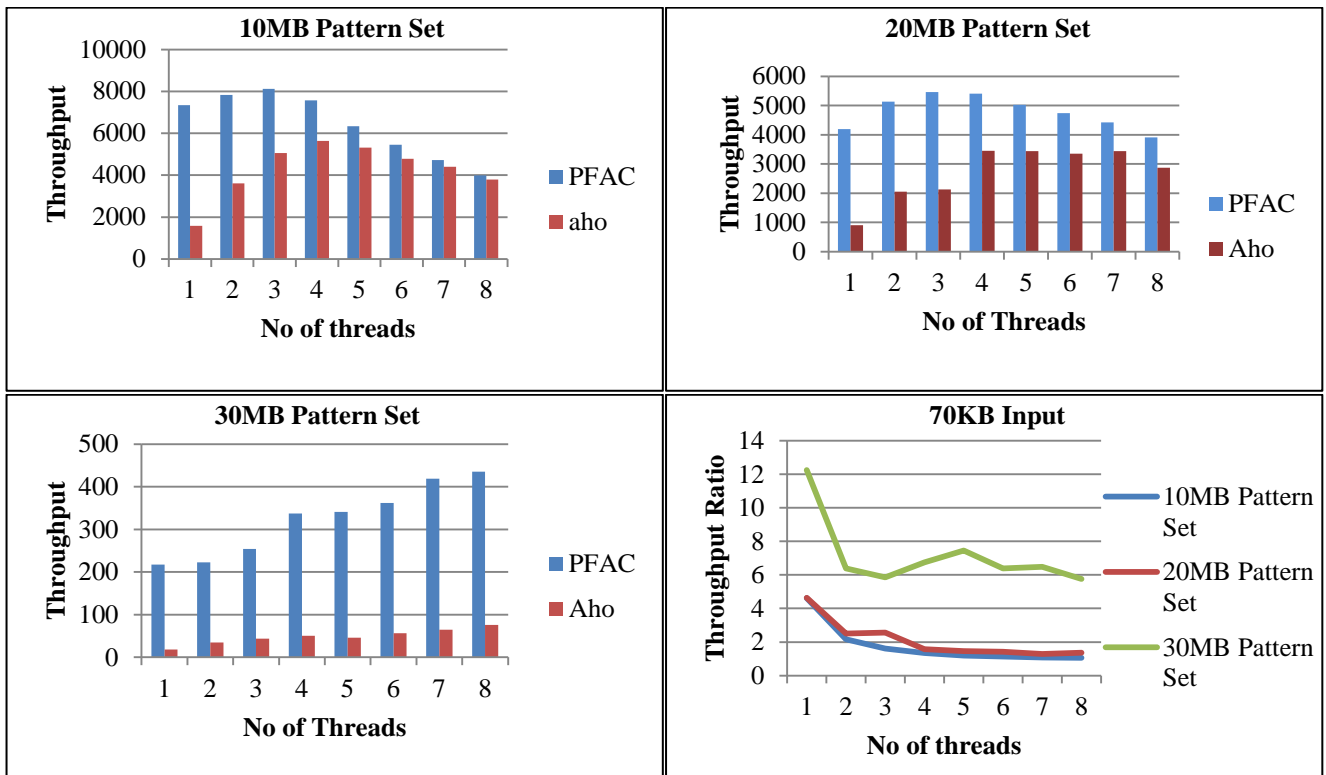

Fig.7. Variation of throughput with number of threads for different pattern sizes and 70KB Input.

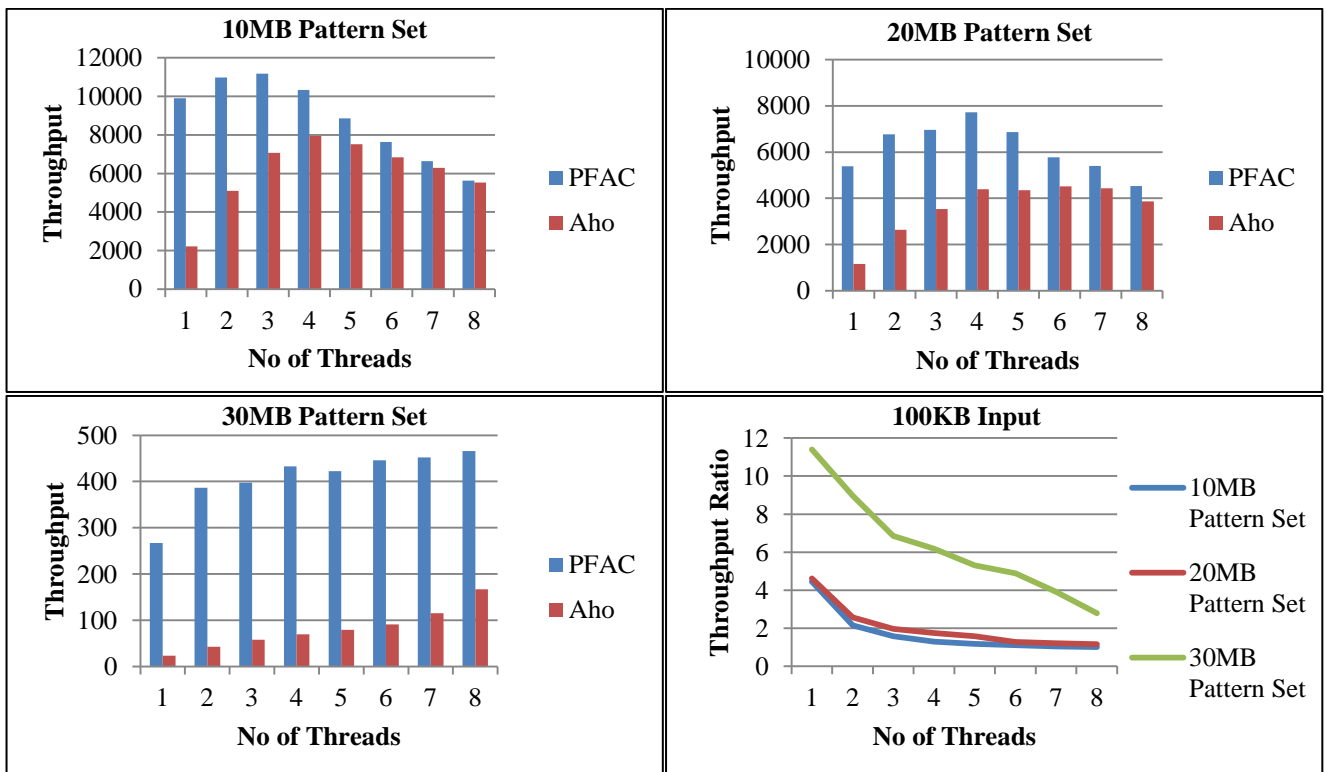

Fig.8. Variation of throughput with number of threads for different pattern sizes and 100KB input.

For considering the scalability of our approach, the experiments were carried out on our implementation and existing implementation [2] by changing the number of patterns (half a million (10MB), one million (20MB) and one and a half million (30MB)) for different input sizes (70KB and 100KB) on a machine with AMD Opteron™ Processor that has 8 core with hyper threading and 8GB of RAM. Fig. 7 shows the variations of throughput of both the implementations (ours and [2]) for different number of threads for 70KB input string as the number of patterns was varied. In the figure, graph 4 shows the throughput ratio between our implementation and the existing implementation [2] for each pattern-set size (10MB, 20MB and 30MB). Fig. 8 depicts the same variable for 100KB input string.

From the results, it can be seen that our method is performing better (with higher throughput) compared to the existing solution [2]. For example, in Fig. 7, consider the graph for the 10MB pattern set. For one thread, our implementation has shown a throughput of 7337 Bps and existing implementation [2] has shown a throughput of 1587 Bps. In addition, the throughput gain gets better with the increasing pattern size for any given input string size. For example, consider graph 4 in the same figure. For one thread, the throughput ratio for 10MB pattern is 4.62 (7337 is divided by 1587) and for 30MB pattern is 12.25. The throughput ratio is increasing with the increasing pattern size. This is because we have reduced the machine constructing time by using the failure less implementation.

## VI. CONCLUSION

Large amount of data in the form of strings is needed to be handled in bio-computing applications. Multiple string matching is an integral part of such applications. Therefore, the performance of multiple string matching algorithms has to be improved. In this paper, we have presented a methodology that uses a multicore processor to achieve improved performance of the Aho-Corasick multiple pattern matching algorithm through parallel manipulation of pattern-sets using POSIX thread utility. With the help of the results and evaluation of our approach, we conclude that our implementation of the Aho-Corasick algorithm is performing better compared to a state of the art parallel implementation of the same algorithm.